\documentclass[structabstract]{aa}  
\usepackage{natbib}
\bibpunct{(}{)}{;}{a}{}{,} 
\usepackage{graphicx}
\usepackage{txfonts}
\usepackage{lscape}

\newcommand{\swift}{{\sl Swift}}
\newcommand{\chandra}{{\sl Chandra}}
\newcommand{\xmm}{{\sl XMM-Newton}}

\def\nh{\hbox{$N_{\rm H}$}}

\begin{document}
   \title{Observations of the post shock break-out emission of SN~2011dh with
\xmm\thanks{Based on observations obtained with \xmm, an ESA science mission with instruments and contributions directly funded by ESA Member States and NASA.}}

   \subtitle{}

   \titlerunning{Post shock break-out emission of SN~2011dh}

   \author{Manami Sasaki
          \inst{1}
          \and
          Lorenzo Ducci\inst{1} 
          }

   \institute{Institut f\"ur Astronomie und Astrophysik, 
              Universit\"at T\"ubingen,
              Sand 1, 
              D-72076 T\"ubingen, Germany,
              \email{sasaki@astro.uni-tuebingen.de}
             }

   \date{Received August 10, 2012; accepted September 16, 2012.}

 
  \abstract
   {After the occurrence of the type cIIb SN~2011dh in the nearby spiral galaxy 
M~51 numerous observations were performed with different telescopes in 
various bands ranging from radio to $\gamma$-rays.}
   {We analysed the \xmm\ and \swift\ observations taken 3 to 30 days after the
SN explosion to study the X-ray spectrum of SN~2011dh. }
   {We extracted spectra from the \xmm\ observations, which took place $\sim$7 
and 11 days after the SN. In addition, we created integrated \swift/XRT spectra 
of 3 to 10 days and 11 to 30 days.}
   {The spectra are well fitted with a power-law spectrum absorbed with 
Galactic foreground absorption. In addition, we find a harder spectral 
component in the first \xmm\ spectrum taken at $t \approx 7$~d.
This component is also detected in the first \swift\ spectrum of
$t = 3 - 10$~d.
} 
   {While the persistent power-law component can be explained as inverse
Compton emission from radio synchrotron emitting electrons, the harder
component is most likely bremsstrahlung emission from the shocked stellar 
wind. Therefore, the harder X-ray emission that fades away after 
$t \approx 10$~d can be interpreted as emission 
from the shocked circumstellar wind 
of SN~2011dh.}


   \maketitle
%

\section{Introduction}

Massive stars at the end of their lives have undergone phases of more or less 
strong mass loss. The evolution of the stars and thus also the way their 
supernovae (SNe) 
evolve, depend strongly on their mass loss rates. The interaction of the SN 
shock wave with the stellar wind creates a hot region around the SN site, in 
which electrons are accelerated and produce radio synchrotron emission. The
hot plasma around the SNe can be observed in X-rays.
Strong X-ray emission is in particular expected when the SN shock wave breaks
out of the star and starts propagating into the circumstellar material
\citep[][and references therein]{2011ApJ...729L...6C,2012ApJ...747L..17C}.
Owing to prompt observations with the \swift\ or the \chandra\ telescope,
X-rays from the interaction of the SN shock with the circumstellar matter
have been detected for a number of SNe \citep[e.g., SN~2006jc or SN~2010jl,]
[respectively]{2008ApJ...674L..85I,2012ApJ...750L...2C}. 
\citet{2008Natur.453..469S} detected a transient X-ray source with \swift,
which was then identified as SN~2008D in the galaxy NGC~2770. 
The X-ray outburst was ascribed to the shock break-out of the SN. 

On May 31, 2011, a supernova explosion was observed 
\citep{2011ATel.3398....1S} in the nearby galaxy M\,51 located at a distance 
of 8.4$\pm$0.7~Mpc \citep{2012A&A...540A..93V}. It was classified as
a type IIb SN \citep{2011ATel.3413....1A}. Several radio, optical, and
X-ray observations followed to study the SN evolution, which also allowed to 
constrain the nature of the progenitor suggesting a compact progenitor star
with a radius of $\sim10^{11}$~cm and thus the classification of SN~2011dh as
a type cIIb SN \citep{2011ApJ...742L..18A,2012ApJ...752...78S}.

\citet{2012ApJ...752...78S} studied SN~2011dh using \swift\ and \chandra\ data
in X-rays and the Submillimeter Array, the Combined Array for Research in 
Millimeter-wave Astronomy, and the Expanded Very Large Array in radio in the
following weeks. They showed that the observed radio emission is synchrotron 
radiation of electrons, accelerated in the forward shock of the SN explosion, 
while inverse Compton (IC) scattering of these electrons produced X-rays. 
They estimated that the break-out 
of the shock out of the compact progenitor must have occured at 
$R_{\rm br} \approx 4 \times 10^{11}$~cm with a rise time for the
break-out pulse of $t_{\rm br} \approx$ 1~min. However, the shock break-out
pulse was not detected by any X-ray or $\gamma$-ray observatory.

We report the detection of a hard X-ray component in the early spectra
of SN~2011dh observed with \xmm\ and \swift. The comparison of the spectra
taken at different times after the SN event ranging from $t$ = 3 to 30~d
allow us to consider this component as 
emission from the interaction of the SN shock with the circumstellar material 
in the immediate surroundings of the progenitor star.

\section{Data}

\subsection{\xmm\ data}

\begin{figure}
\centering     
\includegraphics[width=0.24\textwidth,clip=]{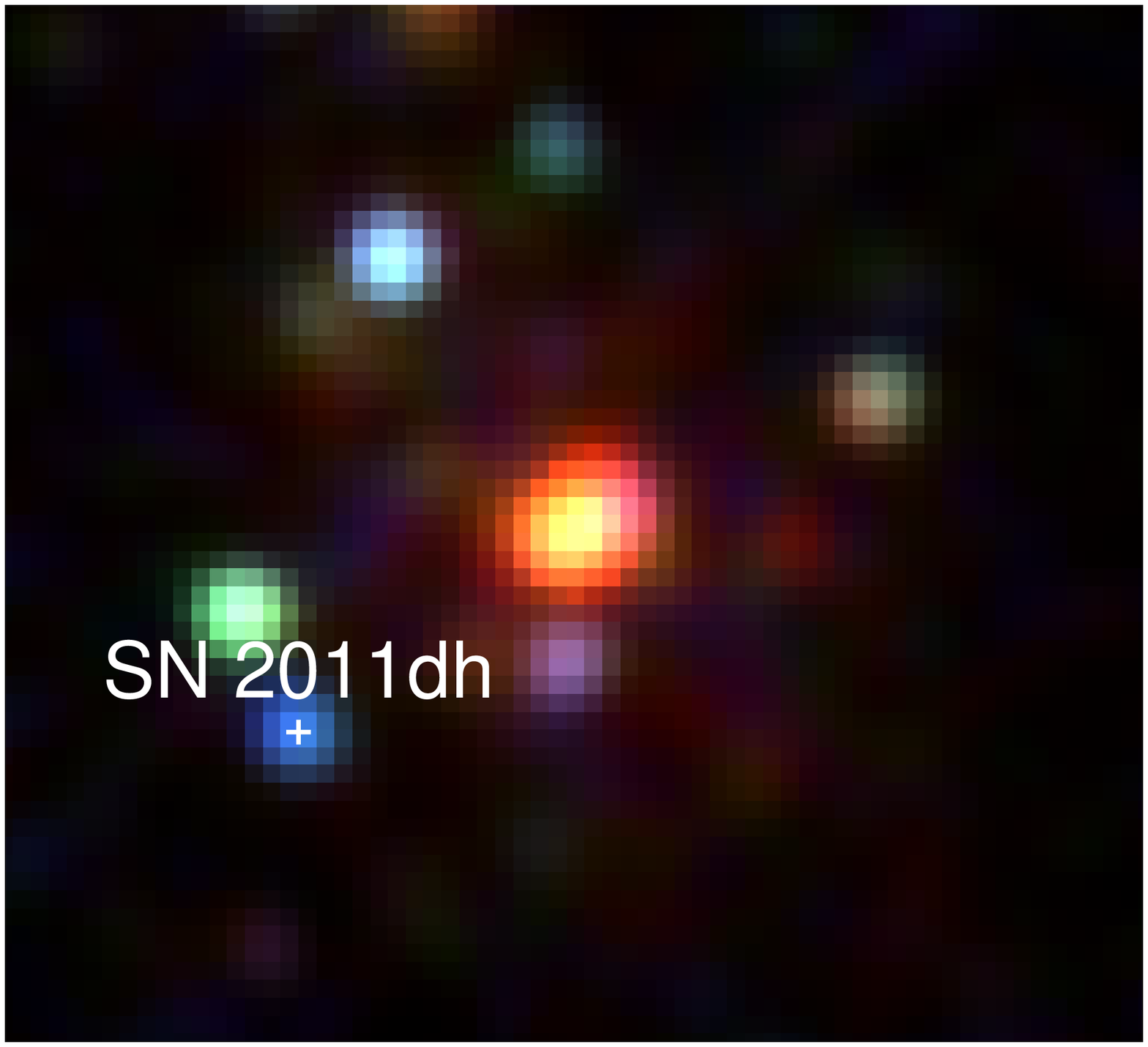}
\includegraphics[width=0.24\textwidth,clip=]{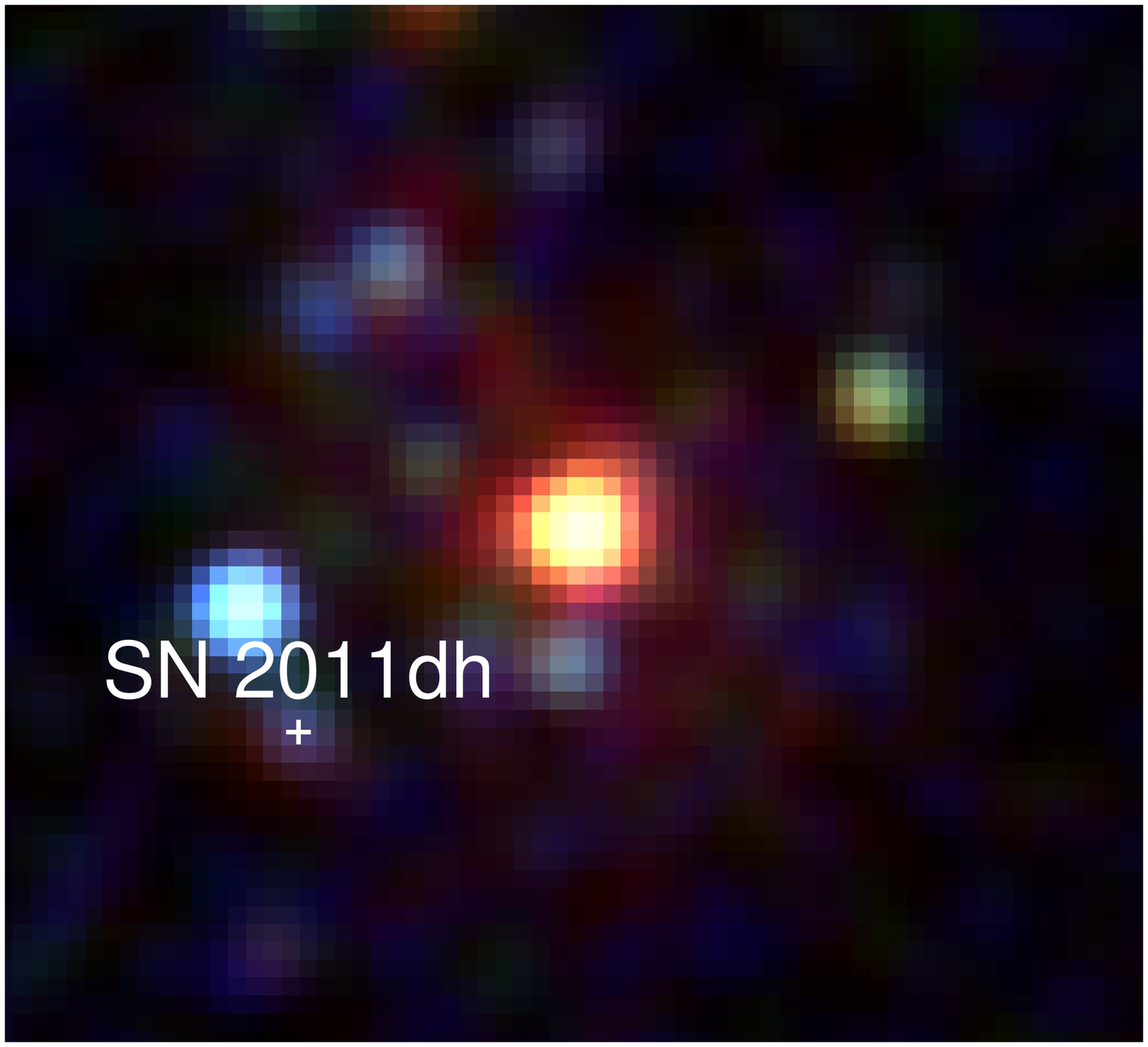}
\caption{
\xmm\ EPIC-pn images $\sim$7~d ({\it left}) and $\sim$11~d ({\it right})
after the SN explosion in the colours
red (0.3 -- 1.0~keV), green (1.0 -- 2.0~keV), and blue (2.0 -- 8.0~keV).
}
\label{xmmima}
\end{figure}

Shortly after SN~2011dh occured, two observations with 
\xmm\ were initiated. The first observation with the ObsID 0677980701 was 
performed at $t \approx$ 7~d after the SN explosion from 
2011/06/07, 5:19:56 to 2011/06/07, 8:30:40 (UTC) with an exposure
of $\sim11$~ks.
All European Photon Imaging Cameras 
\citep[EPICs,][]{2001A&A...365L..18S,2001A&A...365L..27T} 
were operated in full-frame mode and used with the thin filter.
The next observation took place from 2011/06/11, 5:05:03 to 2011/06/11, 
8:15:38 (UTC) at $t \approx$ 11~d with $\sim10$~ks exposure.
The EPICs were used in the same configuration as in the observations before.
After filtering good time intervals, only 4 and 2.6~ks of useful data 
remained for the observations 0677980701 and 0677980801, respectively.

Due to low statistics, the MOS1/2 data are not useful for further analysis.
Therefore, we created images from the EPIC-pn data in the bands 
0.3 -- 1.0~keV, 1.0 -- 2.0~keV, and 2.0 -- 8.0~keV
(Fig.\,\ref{xmmima}). 
As can be seen in the EPIC-pn images, the X-ray source at the position
of SN~2011dh is brighter on 2011/06/07 than on 2011/06/11 and also appears
more blueish. The red-orange appearing source in the center is the nucleus
of M~51 together with an unresolved ultra-luminous X-ray source (ULX).
The other brighter blue or green sources are also ULXs in M~51
\citep{2005ApJ...635..198D}.

We extracted EPIC-pn spectra of SN~2011dh in a circular region with a radius 
of 20\arcsec\ around the optical position. The background spectrum was 
extracted in a 30\arcsec\ radius circle
close to the source, where no X-ray source was detected.
Hereafter, we call the EPIC-pn spectrum of 2011/06/07 spectrum XMM1 and 
that of 2011/06/11 spectrum XMM2.

\subsection{\swift\ data}

SN~2011dh was also observed with the \swift\ satellite in a large number
of observations to study the evolution of the X-ray emission.
\citet{2012ApJ...752...78S} derived the X-ray fluxes of SN~2011dh 
from \swift\ observations taken with the X-ray telescope (XRT) from 
2011/06/03 to 29 and analysed the integrated spectrum from the observations 
of 2011/06/03 to 17. Since we would like to study the \swift\ spectra at 
times comparable to the \xmm\ observations, we collected XRT observations 
of SN\,2011dh
starting $\sim 3$\,d after the SN explosion, in the time period from 
2011/06/03 12:04:02 (UT) to 2011/06/10 17:08:32 (total exposure time
of $\sim 44$\,ks)
and in a later period from 2011/06/11 04:55:01 to 2011/06/30 12:33:50
(total exposure time of $\sim 85$\,ks).
The XRT data obtained in photon-counting (PC) mode were processed
with the standard procedures \citep[XRTPIPELINE v.0.12.6,][]{Burrows05}.
Standard grade filtering (0 -- 12) and screening criteria were applied.

Events for the spectral analysis were accumulated within the same 
circular regions as for the \xmm\ spectra, i.e., with 20\arcsec\ radius 
centered on the optical position of the SN for the source spectrum
and from a source-free circular region of radius 30\arcsec\ close to the SN
for the background.
We used version v.013 of the response matrices
in HEASARC calibration database (CALDB) and
the corresponding ancillary response files created using the task
XRTMKARF.
We call the integrated spectrum of 2011/06/03 to 10 spectrum Swift1 and 
that of 2011/06/11 to 30 spectrum Swift2.

\section{Spectral analysis}

The spectrum Swift1 is merged from the data of 2011/06/03 to 10 and thus 
represents an average spectrum of before 2011/06/10. Similarly,
spectrum Swift2 is an average spectrum of after 2011/06/11. Therefore,
the four analysed spectra have the following chronological order: Swift1, 
XMM1, XMM2, and Swift2, with Swift1 and XMM1 corresponding to spectra
taken at similar times.

\subsection{The longer persistent soft X-ray emission}

We analysed the two \xmm\ and two \swift\ spectra simultaneously to search 
for changes in the spectral components. 
\citet{2011ATel.3456....1P} reported that the \chandra\ spectrum taken 
on 2011/06/12 can be fitted with a power-law spectrum with a photon index
of $\Gamma = 1.4\pm0.3$, absorbed by the Galactic foreground \nh\ 
$= 1.8 \times 10^{20}$~cm$^{-2}$. 
For the \swift\ spectra from 2011/06/03 to 17,  \citet{2012ApJ...752...78S}
determined a photon index of $\Gamma$ = 0.9 -- 1.8.
We therefore fitted the spectra first with a single power-law model, 
also assuming foreground absorption by Galactic \nh\ $= 1.8 \times
10^{20}$~cm$^{-2}$.
The first two spectra Swift1 and XMM1 can be fitted with a lower photon 
index of $\Gamma = 1.1\ (1.0 - 1.3)\footnote{All errors in this paper 
given in brackets are 90\% confidence ranges.}$
with red.\ $\chi^2$ = 1.1 and d.o.f.\ = 33,
whereas the spectra XMM2 and Swift2 
taken after 2011/06/11 are fitted well with $\Gamma = 1.8\ (1.5 - 2.0)$
with red.\ $\chi^2$ = 1.1 and d.o.f.\ = 23.
Therefore, the spectra taken on the first $\sim$10 days after the SN is 
significantly  different than the spectra thereafter.

We also fitted all spectra with a single thermal bremsstrahlung model.
While the temperatures fitted for the first two spectra ($t <$ 10~d)
are unconstrained ($kT_{\rm brems,Swift1} > 9$~keV and 
$kT_{\rm brems,XMM1} > 22$~keV, respectively), the two later spectra
($t >$ 11~d) are both fitted well with temperatures of 
$kT_{\rm brems,XMM2} = 3 (1 - 18)$~keV and 
$kT_{\rm brems,Swift2} = 3 (2 - 8)$~keV (red.\ $\chi^2$ = 1.1 
at d.o.f.\ = 56).

\subsection{The fading hard X-ray emission}

\begin{figure}
\centering     
\includegraphics[height=0.48\textwidth,angle=270,bb=65 34 573 704,clip=]{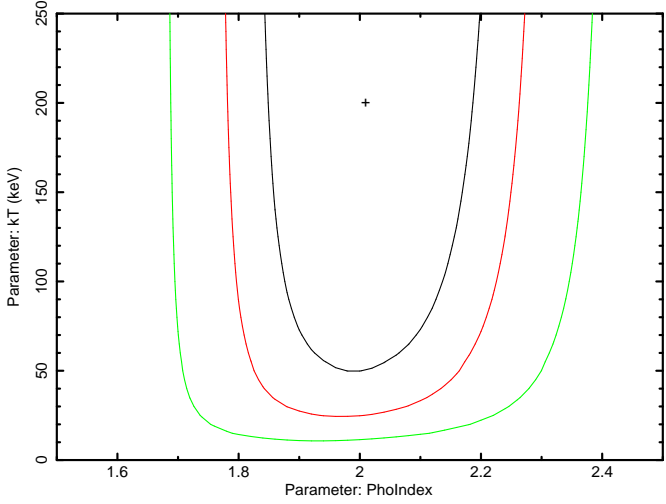}
\caption{
Confidence level contours (63, 90, and 99\%) for the parameters 
$kT_{\rm brems}$ and $\Gamma$.
%
}
\label{contours}
\end{figure}

\begin{figure}
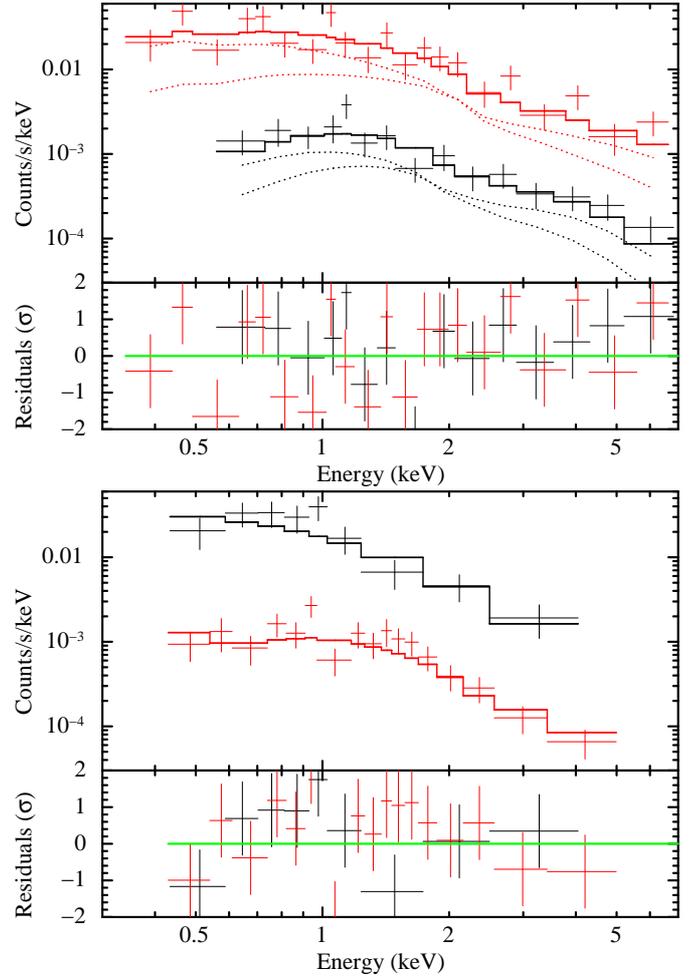

\centering     
\includegraphics[height=0.48\textwidth,angle=270,bb=80 34 568 704,clip=]{brems+pow_12.ps}
\includegraphics[height=0.48\textwidth,angle=270,bb=80 34 568 704,clip=]{pow_34.ps}
\caption{
X-ray spectra of SN~2011dh: stacked \swift\ spectrum for $t < 10$~d 
(Swift1, black in the {\it upper} diagram), \xmm\ EPIC-pn spectrum taken at
$t \approx 7$~d (XMM1, red in the {\it upper} diagram), 
\xmm\ EPIC-pn spectrum taken at $t \approx 11$~d (XMM2, black in the 
{\it lower} diagram), and stacked \swift\ spectrum for $11 < t < 30$~d 
(Swift2, red in the {\it lower} diagram). The spectra in the {\it upper}
diagram are fitted with a combined power-law + bremsstrahlung spectrum, while
the spectra in the {\it lower} diagram are fitted with a power-law spectrum 
only. The \swift\ spectra appear at lower countrate per energy bin due to lower 
effective area of the detector.
}
\label{spectra}
\end{figure}

A likely origin of the additional harder component in the earlier spectra 
is free-free emission from the shocked circumstellar wind
\citep{2003LNP...598..171C}.
If we keep the power-law component and 
include an additional free-free emission component for the earlier 
spectra Swift1 and XMM1, the photon indices of their power-law component
become higher with $\Gamma = 2.0\ (1.6 - 2.5)$
(red.\ $\chi^2$ = 1.1 at d.o.f.\ = 32). 
Moreover, we obtain an additional absorbing column density of 
$\nh_{\rm intr}$  = 7.2\ (1.3 - 15.0) $\times 10^{20}$~cm$^{-2}$.
The flux of the power-law component is the same for all four spectra
with the unabsorbed flux being $F_{\rm pow} (0.3 - 8.0~\mathrm{keV}) 
= 1.1 (0.8 - 1.5) \times 10^{-13}$~erg~cm$^{-2}$~s$^{-1}$.
The temperature of the bremsstrahlung component is not well constrained and 
we can only determine a lower limit of $kT_{\rm brems} > 42$~keV. 
Figure \ref{contours} shows the confidence contours for 
the parameters $kT_{\rm brems}$ and $\Gamma$. The unabsorbed flux of
the bremsstrahlung component is 
$F_{\rm brems} (0.3 - 8.0~\mathrm{keV}) 
= 1.3 (0.9 - 1.7) \times 10^{-13}$~erg~cm$^{-2}$~s$^{-1}$ and
$1.1 (0.7 - 1.6) \times 10^{-13}$~erg~cm$^{-2}$~s$^{-1}$ for 
Swift1 and XMM1, i.e., for $t < 10$~d and $t = 7$~d, respectively.
%
%
%
While this component has disappeared at $t = 11$~d
with an upper limit of the unabsorbed flux of 
$F_{\rm brems} (0.3 - 8.0~\mathrm{keV}) 
< 0.6 \times 10^{-13}$~erg~cm$^{-2}$~s$^{-1}$, the power-law 
component seems to represent emission that is persistent up to 30~d or 
longer.
Figure \ref{spectra} shows the first two spectra
Swift1 and XMM1 with a model consisting of a bremsstrahlung and a
power-law component (upper diagram), while the later spectra XMM2 and Swift2 
are shown with the single power-law fits (lower diagram).

\citet{2003LNP...598...91I} on the other hand discuss the presence
of a blackbody continuum of the shocked hot circumstellar gas.
If we fit the spectra Swift1 and XMM1 with a blackbody model for the 
additional hard component we obtain a temperature of 
$kT_{\rm BB} = 1.6  (1.7-2.8)$~keV. 
%
The luminosities are 
$L_{\rm BB}(\mathrm{<10~d}) = 1.6 (0.8 - 3.7) \times 10^{39}$~erg~s$^{-1}$
and
$L_{\rm BB}(\mathrm{7~d}) = 1.6 (0.8 - 3.1) \times 10^{39}$~erg~s$^{-1}$
for Swift1 and XMM1, respectively.
From the fits of the spectra XMM2 and Swift2, we obtain the following
upper limits:
$L_{\rm BB}(\mathrm{11~d}) < 0.8 \times 10^{39}$~erg~s$^{-1}$
and for XMM1,
$L_{\rm BB}(\mathrm{11-30~d}) < 0.3 \times 10^{39}$~erg~s$^{-1}$.

\section{Discussion}

Many SNe show a softening of their X-ray spectrum days to months
after the SN event 
\citep[e.g.,][and references therein]{2005ApJ...632L..99I}.
If the X-ray emission is mainly due to IC process as proposed for SN~2011dh
by \citet{2012ApJ...752...78S}, one expects $dL_{\rm X}/dE \propto\ 
E^{-(p-1)/2}$, with $p$ being the spectral index of the energy of the 
injected electrons responsible also for the radio synchrotron emission 
\citep[e.g.,][]{2006ApJ...641.1029C}. 
This, of course, is based on a simple assumption that the particle spectrum
does not change over a broad spectral range. In reality, some acceleration 
processes or particle losses can modify the spectrum. 
As the power-law fits of XMM1 and Swift1 vs.\ Swift2 and XMM2 have shown
the photon indices are significantly different ($\Gamma$ = 1.1 [1.0 -- 1.3]
and 1.8 [1.5 -- 2.0], respectively) and indicate temporal change in the X-ray
spectrum.
Assuming that the particle spectrum does not change its shape significantly,
the softening of the X-ray spectrum is most likely not caused by the change of 
the slope of the non-thermal spectrum, but rather by the change of the emission 
components. 
The X-ray emitting gas behind the blast wave of the SN shock is
expected to have high temperatures of ~100 keV or higher, while the gas 
behind the reverse shock is cooler 
\citep[$\sim1$ -- 10~keV][]{2003IAUS..214..113I}. Hence a 
likely explanation for the softening of the X-ray spectrum is that
the reverse shock emission becomes more dominant.

The analysis of the \xmm\ EPIC-pn and \swift\ XRT spectra of SN~2011dh have 
shown that the X-ray spectrum changes during $\sim$10~d after the SN event.
If we assume that the longer persistent X-ray emission component 
can be interpreted as thermal bremsstrahlung, we obtain a temperature of 
$\sim$3~keV after $t \approx$11~d, i.e., $\sim3 \times 10^7$~K, corresponding 
to a shock velocity of $\sim1500$~km~s$^{-1}$, which is far lower 
than what was measured in the optical spectra at these times 
\citep[$\sim$10000~km~s$^{-1}$,][]{2011ApJ...742L..18A}.
Therefore, this component is better to be identified as reverse shock
emission or IC emission.
In this case, the additional hot component observed only until $\sim$10~d 
after the SN event might be emission from circumstellar gas shocked by the 
forward shock.

A similar X-ray emission was also detected for SN~1993J with ROSAT few days 
after the SN explosion, which occured in the galaxy M~81 at a distance of 
3.6~kpc \citep{1994Natur.367..621Z}. This emission was fitted well with a 
power-law spectrum with $\Gamma$ = 1.0$\pm$0.25 or a bremsstrahlung
spectrum with $kT_{\rm brems} > 7$~keV, and decreased exponentially
during the $\sim$40~d in which it was observed.

Assuming free-free emission for the hard component, we
obtain a lower limit for the temperature of $kT_{\rm brems} > 42$~keV.
The flux at $t = 7$~d is $F_{\rm brems} (0.3 - 8.0~\mathrm{keV}) 
= (1.1 \pm\ 0.5) \times 10^{-13}$~erg~cm$^{-2}$~s$^{-1}$, corresponding
to $L_{\rm brems} (0.3 - 8.0~\mathrm{keV}) 
\approx 9.3 \times 10^{38}$~erg~s$^{-1}$. 
%
%
%
Using the lower limit for the temperature, we obtain a lower limit for 
the shock velocity of
\begin{equation}
v_{\rm s} = \sqrt{\frac{16 kT_{\rm brems}}{3 \bar{m}}}
> 9800$~km~s$^{-1},
\end{equation}
with a mean mass per free particle for fully ionised 
plasma of $\bar{m} = 0.61~m_{\rm p}$. \citet{2012ApJ...752...78S}
estimated $v_{\rm s} \approx 0.1~c$, which is about three times 
higher than this lower limit. The \xmm\ and \swift\ spectra that
only extend up to $\sim$10~keV do not allow to constrain the temperature
for a bremsstrahlung model. 
Using  $v_{\rm s} \approx 0.1~c$ 
derived by \citet{2012ApJ...752...78S}, we estimate that
at $t = 7$~d, the shock must have reached a radius of
$R_{\rm s} \approx v_{\rm s} t \approx 1.8 \times 10^{15}$~cm.
%
%
Therefore, from the bremsstrahlung spectrum at $t = 7$~d, we derive a 
down-stream density of $n$ = 6.4 $\times 10^7$~cm$^{-3}$ at 
$R_{\rm s} \approx 1.8 \times 10^{15}$~cm.

On the other hand, we can also estimate the stellar wind density based on 
assumption of the progenitor star. SN~2011dh was classified as a type IIb SN 
with a compact progenitor, making a Wolf-Rayet (WR) star a likely progenitor 
candidate \citep{2010ApJ...711L..40C}. 
The number density of the wind at a distance $r$ from the WR star is given by 
the continuity equation:
\begin{equation} \label{cont eq}
n(r) = \frac{\dot{M}}{4 \pi \mu m_{\rm p} r^2 v(r)}
\end{equation}
where $\dot{M}$ is the mass-loss rate, $\mu$ is the mean atomic weight of the 
wind material, and $v(r)$ is the $\beta$-velocity law 
\citep{1975ApJ...195..157C}:
\begin{equation} \label{eq velocity}
v(r) = v_\infty \left ( 1 - \frac{bR_*}{r} \right )^\beta.
\end{equation}
$R_*$ is the stellar radius, $v_\infty$ the terminal velocity, 
$b=0.9983$ a parameter that 
ensures that $v(R_*) > 0$, and $\beta=4-8$ for WR stars 
\citep{1997A&A...321..268S}.
The volume integral of equation (\ref{cont eq}) over a shell with radii
$R_{\rm br}$ and $R_{\rm s}$ centered at the position of the WR star gives 
the total number of particles:
\begin{equation} \label{equa N}
N= \int_V n(r)dr \mbox{.}
\end{equation}
Thus, from equation (\ref{equa N}), it is possible to obtain the average 
number density in that shell:
\begin{equation} \label{n average}
\bar{n} = N \times \left[ \frac{4}{3}\pi (R_{\rm s}^3 -R_{\rm br}^3) 
\right]^{-1} \mbox{.}
\end{equation}
\citet{2012ApJ...752...78S} estimated a stellar radius of
$R_* = 10^{11}\mbox{\,cm}$, a shock break-out radius of
$R_{\rm br} = 4 \times 10^{11}\mbox{\,cm}$, 
and a mass loss rate of
$\dot{M} = 3 \times 10^{-5}\mbox{\,M}_\odot\mbox{\,yr}^{-1}$.
With a typical wind velocity of
$v_\infty = 10^3\mbox{\,km\,s}^{-1}$ and
$R_{\rm s} = 1.8 \times 10^{15}\mbox{\,cm}$ at t = 7~d,
we obtain a mean stellar wind density of
$\bar{n} \approx 8\times 10^5\mbox{\,cm}^{-3}$ from equation (\ref{n average}).
This calculation is based on the assumption, that the stellar wind density is
highest close to the stellar surface and decreases with $\sim r^{-2}$.
However, the stellar wind material might as well form a dense shell around
the star at a certain distance, as was most likely the case for 
SN~2006jc \citep{2008ApJ...674L..85I}. This SN showed an increase in X-rays 
about 100~days after the explosion which was interpreted as emission from
shock heated shell at $R_{\rm s} \approx 10^{16}$~cm with a thickness
of $\Delta R \approx 2 \times 10^{15}$~cm.
A similar condition might have been the case also for SN~2011dh,
however, with a shell being located closer to the star, which resulted in 
an earlier rise and decay of the X-ray emission.

\section{Summary}

Two \xmm\ observations of SN~2011dh performed $\sim$7 and 11 days after the 
discovery revealed that 
the X-ray spectrum changes significantly at about 10 days after the explosion. 
The analysis of the \xmm\ EPIC-pn spectra and additional 
stacked
\swift/XRT spectra 
extracted from many observations of
similar periods supports the existence of two spectral components with 
the harder component disappearing after $\sim$10 days. 
The softer component can be identified either as IC emission as suggested by 
\citet{2012ApJ...752...78S} or as reverse shock emission with an unabsorbed 
flux of $F_{\rm soft} (0.3 - 8.0~\mathrm{keV}) 
\approx 1 \times 10^{-13}$~erg~cm$^{-2}$~s$^{-1}$ up to $t \approx 30$~d.
The flux of the hard component if we assume that it is bremsstrahlung emission,
is $F_{\rm brems} (0.3 - 8.0~\mathrm{keV}) 
= (1.1 \pm\ 0.5) \times 10^{-13}$~erg~cm$^{-2}$~s$^{-1}$ at $t = 7$~d, and 
thus comparable to the soft component, and decreases quickly thereafter.
This indicates that this emission has its origin in the circumstellar matter
that extends to $\sim\ 10^{15}$~cm and was heated by the SN shock.

\begin{acknowledgements}
This research has made use of data obtained from the High Energy Astrophysics 
Science Archive Research Center (HEASARC), provided by NASA's Goddard Space 
Flight Center.
This work was support by the Deutsche Forschungsgemeinschaft through the Emmy 
Noether Research Grant SA 2131/1.
\end{acknowledgements}

\bibliographystyle{aa} 
\bibliography{../../bibtex/xraytel,../../bibtex/sne,../../bibtex/nearbygal}


\end{document}